# Extracellular electrical signals in a neuron-surface junction: model of heterogeneous membrane conductivity

Pavel M. Bulai (pavel.bulai@gmail.com),
Pavel G. Molchanov, Andrey A. Denisov,
Taras N. Pitlik, Sergey N. Cherenkevich

Department of Biophysics, Physics Faculty, Belarusian State University
Nezavisimosty Av. 4, 220030 Minsk, Belarus

**Abstract**
Signals recorded from neurons with extracellular planar sensors have a wide range of waveforms and amplitudes. This variety is a result of different physical conditions affecting the ion currents through a cellular membrane. The transmembrane currents are often considered by macroscopic membrane models as essentially a homogeneous process. However, this assumption is doubtful, since ions move through ion channels, which are scattered within the membrane. Accounting for this fact, the present work proposes a theoretical model of heterogeneous membrane conductivity. The model is based on the hypothesis that both potential and charge are distributed inhomogeneously on the membrane surface, concentrated near channel pores, as the direct consequence of the inhomogeneous transmembrane current. A system of continuity equations having non-stationary and quasi-stationary forms expresses this fact mathematically. The present work performs mathematical analysis of the proposed equations, following by the synthesis of the equivalent electric element of a heterogeneous membrane current. This element is further used to construct a model of the cell-surface electric junction in a form of the equivalent electrical circuit. After that a study of how the heterogeneous membrane conductivity affects parameters of the extracellular electrical signal is performed. As the result it was found that variation of the passive characteristics of the cell-surface junction, conductivity of the cleft and the cleft height, could lead to different shapes of the extracellular signals.

## Introduction

Techniques of the extracellular electrical recording and stimulation made a significant progress since an introduction of a first planar microelectrode arrays and field-effected transistors (Thomas *et al.*, 1972; Bergveld *et al.*, 1976; Gross *et al.*, 1977) (Fig. 1). Microelectrode arrays fabricated according to modern semiconductor technologies often integrate multiple elements of passive and active circuitry. Arrays are used to effectively record, amplify and condition extracellular signals as well to perform extracellular stimulation (Eversmann *et al.*, 2003; Lambacher *et al.*, 2004). Nowadays microelectrode arrays are considered as a basic platform for the development of cell-based sensors (Parce *et al.*, 1989; DeBusschere & Kovacs, 2001; Yeung *et al.*, 2001; Pancrazio *et al.*, 2003).

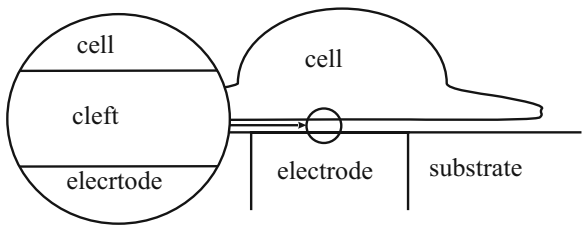

Fig. 1. Schematic view of an electrode covered with neurons

The application of microelectrode arrays gave a start to a long-term investigation of different dynamic processes taking place in cell cultures and tissue slices (Besl & Fromherz, 2002; Heuschkel *et al.*, 2002; Jimbo *et al.*, 2006). A diversity of shapes and a wide range of amplitudes of signals recorded with planar electrodes from different neurons have been reported (Gross, 1979; Regehr *et al.*, 1989; Bove *et al.*, 1995; Breckenridge *et al.*, 1995; Jenkner & Fromherz, 1997; Schatzthauer & Fromherz, 1998; Fromherz, 1999; Ruardij *et al.*, 2009). Signals were generally classified (arranged in types) according to the waveform and amplitude (Fromherz, 2003). This classification is used conventionally for spike detection and sorting in the cell population (Salganicoff *et al.*, 1988; Sarna *et al.*, 1988) as well as for an individual cell characterization (Stett *et al.*, 2003). In cited papers all signal types where explained to originate from and simulated on the basis of several possible mechanisms: the asymmetry of the cell soma and neurites shapes (Bove *et al.*, 1994; Gold *et al.*, 2006), variability of sealing resistance in the neuron-electrode electrical contact (Grattarola & Martinoia, 1993) and the membrane channel distributions (Jenkner & Fromherz, 1997; Schatzthauer & Fromherz, 1998; Fromherz, 1999; Buitenweg *et al.*, 2002).

Up to date main models for a signal simulation are: current sources field integration (Plonsey, 1964; Plonsey & Barr, 2007), equivalent electric circuits (Regehr *et al.*, 1989; Grattarola & Martinoia, 1993) and geometry-based finite-element modeling (Buitenweg *et al.*, 2002, Heuschkel *et al.*, 2002). In these models the membrane current is described with a stationary continuity equation. In the integrated form, the stationary continuity equation corresponds to the Kirchhoff's law. According to the Kirchhoff's law the membrane current is a sum of capacitive and ionic currents:

$$j_m(\psi_m) = c_m \frac{\partial \Delta \psi_m}{\partial t} + j_c(\Delta \psi_m), \quad (1)$$

where, $j_m$ – the total membrane current density, $\psi_m$ – a membrane potential, $\Delta\psi_m$ – a transmembrane potential, $c_m$ – a membrane specific capacitance, $t$ – the time variable, $j_c$ – the ionic current density.

It should be pointed out, that the Kirchhoff's law for the membrane current in Eq. 1 assumes the homogeneous flow of the charge through the membrane. However, on the biological basis, the transmembrane current is flowing through ion channels and not through the whole cellular membrane. Total channels cross-section area is less than 0.01 % of the total membrane area (Nicholls *et al.*, 2001). In addition, the distance between channels of identical types is often can be even larger than the distance between the cellular membrane and the sensor surface (the distance between channels can be estimated from the conductivity of the membrane and the channels).

In the case of a homogeneous charge flow in Eq. 1, the value of the transmembrane current is the function of the membrane conductivity only. However, if the charge is transferred through the membrane channels, then the conductivity of solution near the membrane should directly influences the charge relaxation on membrane surfaces and, consequently, the total membrane current (Fig. 2).

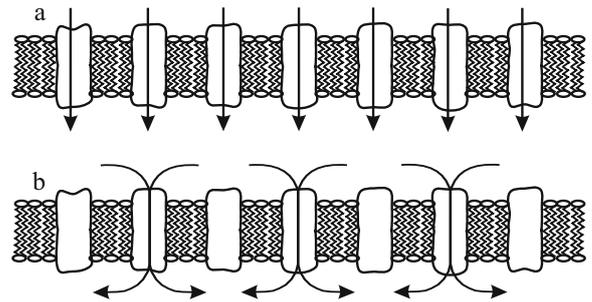

Fig. 2. Difference between homogenous (a) and heterogeneous (b) membrane conductivity

The present work proposes a theoretical model of the heterogeneous membrane conductivity. The model is based on the hypothesis that the electrical potential as well as charge are distributed inhomogeneously on membrane surfaces due to the transmembrane current inhomogeneity. The model is expressed with a system of continuity equations and has non-stationary and quasi-stationary forms.

To obtain parameters for the heterogeneous membrane conductivity model, a charge flow through the single membrane pore (channel) was computationally simulated. Potential-to-charge ratio in the vicinity of the membrane channel was estimated as a function of the membrane capacitance, channel radius and average channel density empirically.

On the next step the equivalent electric element of the cellular membrane was developed. It was based on the heterogeneous membrane conductivity model which implies that the membrane conductivity of the element has dependence on the medium conductivity on both sides of the membrane.

This equivalent electric element of the heterogeneous membrane current was further employed in a model of the cell-surface electric junction. Build in a form of the equivalent electrical circuit the model was used to evaluate effects of the heterogeneity in the membrane conductivity on signal parameters. Main types of extracellular electrical signals have been acquired when this model was subjected to various cell-surface junction heights and junction conductivities.

**Model of the heterogeneous membrane conductivity**

When the charge is traveling in and out the channel pore it creates the region of excess charge $Q_e$ and overpotential $\psi_e$ just near the end of the channel. The values of the additional excess charge $Q_e$ and overpotential $\psi_e$ can be conveniently defined in relation to a spatially homogeneous charge $Q_m$ and potential $\psi_m$ on the rest of the membrane surface. Existence of the overpotential near the pore allows to define a transchannel potential (total local potential over the channel) as the sum $\Delta\psi_m + \Delta\psi_e$, which is obviously different from the simple transmembrane potential $\Delta\psi_m$ and, which is actually should be used when one calculates the conductance of the potential-dependent channel.

Following this, a general scheme of the model can be described as a two-step process: charge transfer into the region with the excess charge near the end of membrane channel following by the immediate drifting of the transferred charge into the spatially homogeneous charge region nearby (Fig. 3).

The rate of change of the excess charge $Q_e$ is a sum of a current through the channel $J_c$ and a lateral relaxation current $J_e$ as depicted in the Fig. 3. At the same time the rate of change of the spatially homogeneous charge $Q_m$ is equal to the sum of the lateral relaxation current $J_e$ and a transmembrane current $J_m$ (the charge migration current). These statements can be written as:

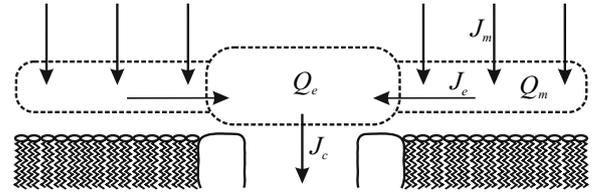

Fig. 3. Scheme of the heterogeneous membrane conductivity model: relative positions of the excess $Q_e$ and homogeneous $Q_m$ charges, directions of the channel $J_c$, lateral $J_e$ and total membrane $J_m$ currents

$$\begin{cases} \dfrac{\partial Q_m}{\partial t} = -J_e + J_m, \\ \dfrac{\partial Q_e}{\partial t} = J_e - J_c. \end{cases} \quad (2)$$

In the system of Eqs. 2 the value of the lateral relaxation current $J_e$ is the total current flowing inward through an imaginary closed surface covering the excess charge $Q_e$ region. Applying the Ohm's law and Gauss-Ostrogradsky theorem to the Gauss's law, the following equation can be derived:

$$J_e = -\dfrac{\delta}{\varepsilon} Q_e, \quad (3)$$

where, $\delta$ – a conductivity of a solution, $\varepsilon$ – the dielectric permittivity of the solution.

Since the value of the current through the channel $J_c$ now is a function of the total potential $\Delta\psi_m + \Delta\psi_e$ and the value of the transmembrane current is a function of the spatially homogeneous potential $\psi_m$ the system of Eqs. 2 may be rewritten in the next form:

$$\begin{cases} \dfrac{\partial Q_m}{\partial t} = \dfrac{\delta}{\varepsilon} Q_e + J_m(\psi_m), \\ \dfrac{\partial Q_e}{\partial t} = -\dfrac{\delta}{\varepsilon} Q_e - J_c(\Delta\psi_m + \Delta\psi_e). \end{cases} \quad (4)$$

In the system of Eqs. 4, values of homogeneous and excess charges relate to the homogeneous potential and overpotential accordingly:

$$\begin{aligned} Q_m &= C_m \Delta\psi_m, \\ Q_e &= C_m K_e \psi_e. \end{aligned} \quad (5)$$

where, $K_e$ – a some coefficient, which connects values of the excess charge and overpotential near the channel end. $K_e$ will be discussed and estimated later.

Substitution of Eqs. 5 into Eqs. 4 gives a non-stationary form of the heterogeneous membrane conductivity model:

$$\begin{cases} C_m \dfrac{\partial \Delta\psi_m}{\partial t} = \dfrac{\delta}{\varepsilon} C_m K_e \psi_e + J_m(\psi_m), \\ C_m K_e \dfrac{\partial \psi_e}{\partial t} = -\dfrac{\delta}{\varepsilon} C_m K_e \psi_e - J_c(\Delta\psi_m + \Delta\psi_e). \end{cases} \quad (6)$$

In the second equation of Eqs. 6 the ratio $\varepsilon / \delta = \tau$ is a time constant, which defines a rate of the excess charge relaxation. In the case of physiological saline estimation gives $\tau \approx 10^{-9}$ sec. It is much less than the channel activation time. As result one may conclude that the excess charge reaches a stationary value much faster than the homogeneous charge and transmembrane potential do. With this condition met the second differential equation can be replaced by the algebraic Eqs. 7.

$$\begin{cases} C_m \dfrac{\partial \Delta \psi_m}{\partial t} = \dfrac{\delta}{\varepsilon} C_m K_e \psi_e + J_m(\psi_m), \\ 0 = -\dfrac{\delta}{\varepsilon} C_m K_e \psi_e - J_c(\Delta \psi_m + \Delta \psi_e). \end{cases} \quad (7)$$

In the system of Eqs. 7 performing addition of the second equation to the first one leads to the equation (first in the Eq. 8) in the form similar to the Eq. 1. Value of the overpotential $\psi_e$ could be derived from the second equation in Eqs. 7. The system of Eqs. 8 is a quasi-stationary form of the heterogeneous membrane conductivity model.

$$\begin{cases} C_m \dfrac{\partial \Delta \psi_m}{\partial t} = -J_c(\Delta \psi_m + \Delta \psi_e) + J_m(\psi_m), \\ \psi_e = -\dfrac{\varepsilon}{\delta} \dfrac{J_c(\Delta \psi_m + \Delta \psi_e)}{C_m K_e}. \end{cases} \quad (8)$$

It can be seen from the second equation in Eqs. 8 that the smaller conductivity of the environment near the channels pore is, the greater overpotential might appear.

Homogeneous membrane conductivity (Eq. 1) is a special case of model of the heterogeneous membrane conductivity (Eqs. 8): the value of overpotential $\psi_e$ becomes insignificant under conditions that conductivity of solution $\delta$, factor $K_e$ are big or/and transchannel current $J_c$ is small.

**Channel density factor**

The coefficient $K_e$ that binds values of the excess charge and overpotential near the channel was introduced in the second equation of Eqs. 5. This coefficient has a natural dependence both on the channel as well patch geometry and dimensions making it hard to be described analytically in general. However, numerical computational simulations of the ion current flowing through the membrane pore (channel) provides a convenient means to obtain this coefficient at least for a specific case.

The geometry that was used in simulations represents a cylinder separated onto two halves with a membrane containing a single pore. The cylinder has a height equal to 400 nm and a radius $r_m$=200 nm. The thickness of the membrane is 10 nm and the radius of the pore is $r_c$=0÷200 nm. The whole geometry has an axial symmetry. The compartment and channel are considered to be filled with the 0.1 M binary aqueous electrolyte (KCl) at 300 K. A relative permittivity of the membrane is equal to 4.

The transient drift-diffusion (Nernst-Planck-Poisson) problem was used to describe the spatio-temporal distribution of potential and charge within the system. Boundary conditions for concentrations considered the insulation barrier at the compartment and membrane surfaces. Boundary conditions which are related to the potential distribution were: absence of any charges on the side, top and bottom surfaces of the geometry, continuity of the electric potential on the membrane and pore boundaries. Potential at a point in the middle of membrane at the compartment side was taken to be a zero reference potential.

The drift-diffusion problem was solved with the finite-element method using a program environment of COMSOL Multiphysics (COMSOL Group). The application modes were chosen to be the "Nernst-Planck without Electroneutrality" and "Electrostatic". Space dimension had 2D axial symmetry. A non-uniform grid with a higher density near the membrane (element size 1 nm) and pore (element size 0.1 nm) was used. Computations were performed with the BDE time depended solver and direct (UMFPACK) linear system solver.

To set up initial conditions the membrane was allowed to be charged by applying a step of the transmembrane potential of 100 mV. The transmembrane potential was applied by setting a fixed potential on the top and bottom of the geometry compartments. At the appropriate time after this the spatially homogeneous charge has appeared near the membrane. After the charging, top and bottom boundaries of the compartment were set to have a zero charge and the membrane started a slow discharge process by ions drifting through the pore. During the drifting phase the electric potential and surface charge density at the plane of the membrane side were observed. The surface charge density was obtained by integration of a spatial charge density in the direction orthogonal to the membrane. The simulation was performed for a set of different radiuses of the pore: from 0 to 200 nm.

Results of the simulation for the channel radius $r_c$=5 nm at an arbitrary selected time (as an example) are shown on the Fig. 4. Corresponding profiles of the potential and surface charge density on the membrane/channel surface are presented on the Fig. 5.

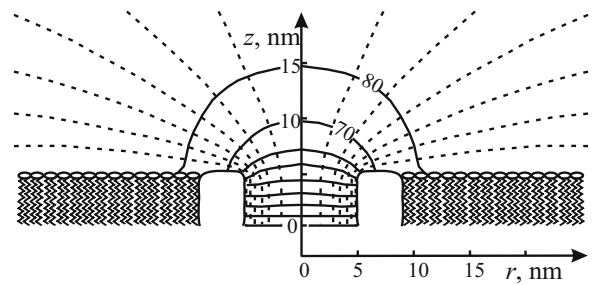

Fig. 4. Results of the solution for the mixed Nernst-Planck-Poisson problem in cylindrical coordinates ($z$, $r$) for the channel of radius $r_c$=5 nm in an arbitrary point of time: equipotential surfaces and current lines

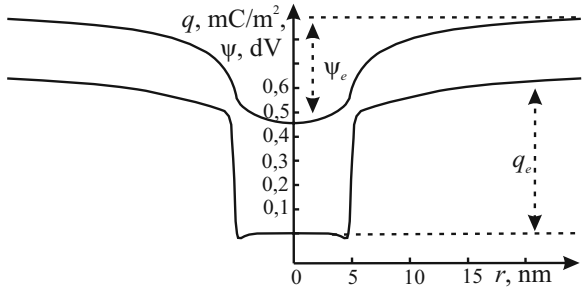

Fig. 5. Results of the solution for the mixed Nernst-Planck-Poisson problem in cylindrical coordinates ($z, r$) for the channel of radius $r_c$=5 nm in an arbitrary point of time: profiles of the potential and surface charge density on the membrane/channel surface

Excess charge density $q_e$ and overpotential $\psi_e$ were obtained in the point that lay on the channel axis just near the channel end. From simulations for different channel radiuses, the relationship between $q_e$ and $\psi_e$ was found to be independent on the channel current, but it did depend on the channel and patch radiuses (Fig. 6). This relationship appeared to be well approximated by the curve given by the Eq. 9, where, $\lambda$ is the Debye length:

$$\frac{q_e}{\psi_e} = c_m \frac{4\lambda + r_c}{r_c} \frac{2r_m - r_c}{r_m - r_c}. \tag{9}$$

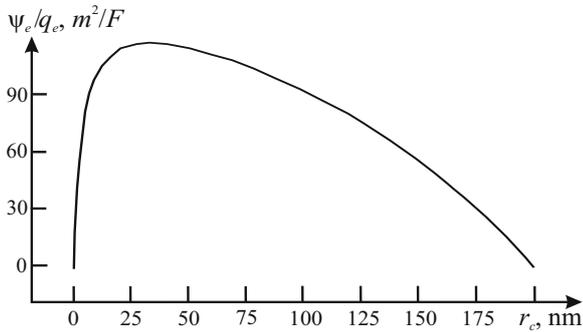

Fig. 6. Results of the solution for the mixed Nernst-Planck-Poisson problem for different channel radiuses: relationship between the excess charge density $q_e$ and overpotential $\psi_e$ in the dependence on channel radius

$K_e$ function is expressed from Eqs. 5 and Eq. 9, using $Q_e = \pi r_c^2 q_e$ and $C_m = \pi r_m^2 c_m$, to get the following formula:

$$K_e = \frac{r_c^2}{r_m^2} \frac{4\lambda + r_c}{r_c} \frac{2r_m - r_c}{r_m - r_c}. \tag{10}$$

Eq 10 can be further simplified in the case when: $r_m \gg r_c$ - which means a low channel density; and $r_c \approx \lambda$ - which means that a minimal size of a screened charge in the solution is equal to the Debye length. By applying these assumptions, one can express:

$$K_e = 10 \frac{\lambda^2}{r_m^2}. \tag{11}$$

After introduction of the ion channel density $\eta = 1/\pi r_m^2$, Eq 11 can be rewritten as:

$$K_e = 10\pi\eta\lambda^2. \tag{12}$$

Thus, $K_e$ function can be called a channel density factor.

**Equivalent electric element of heterogeneous membrane current**

Equations of the heterogeneous membrane conductivity model could be modified into the equivalent electric element of membrane current that can be used in the cell-surface junction point contact model.

The non-stationary form of the heterogeneous membrane conductivity model in Eqs. 6 can be rewritten in form of the following Eq. 13 and Eqs. 14:

$$J_m(\psi_m) = C_m \frac{\partial \Delta\psi_m}{\partial t} + J_e(\psi_m), \tag{13}$$

Where:

$$\begin{cases} J_e(\psi_m) = -\frac{\delta}{\varepsilon} C_m K_e \psi_e, \\ \frac{\partial \psi_e}{\partial t} = -\frac{\delta}{\varepsilon} \psi_e - \frac{J_c(\Delta\psi_m + \Delta\psi_e)}{C_m K_e}. \end{cases} \tag{14}$$

System of Eqs. 14 could be rewritten in terms of the capacitance and current surface density:

$$\begin{cases} j_e(\psi_m) = -\frac{\delta}{\varepsilon} c_m K_e \psi_e, \\ \frac{\partial \psi_e}{\partial t} = -\frac{\delta}{\varepsilon} \psi_e - \frac{j_c(\Delta\psi_m + \Delta\psi_e)}{c_m K_e}. \end{cases} \tag{15}$$

System of Eqs. 15 describes the current density only at one side of the membrane. To complete the membrane description with a second side all equations in the system of Eqs. 15 were doubled for the inner (*in*) and outer (*out*) currents densities:

$$\begin{cases} j_e^{in}(\psi_m^{in}) = -\frac{\delta^{in}}{\varepsilon} c_m K_e \psi_e^{in}, \\ \frac{\partial \psi_e^{in}}{\partial t} = -\frac{\delta^{in}}{\varepsilon} \psi_e^{in} - \frac{j_c(\Delta\psi_m + \Delta\psi_e)}{c_m K_e}, \\ j_e^{out}(\psi_m^{out}) = -\frac{\delta^{out}}{\varepsilon} c_m K_e \psi_e^{out}, \\ \frac{\partial \psi_e^{out}}{\partial t} = -\frac{\delta^{out}}{\varepsilon} \psi_e^{out} + \frac{j_c(\Delta\psi_m + \Delta\psi_e)}{c_m K_e}, \\ \Delta\psi_m = \psi_m^{in} - \psi_m^{out}, \\ \Delta\psi_e = \psi_e^{in} - \psi_e^{out}. \end{cases} \tag{16}$$

Finally, currents were written down for all type of ions, which in our case are Na, K, Cl:

$$\begin{cases} j_e^{in}\left(\psi_m^{in}\right) = -\dfrac{\delta^{in}}{\varepsilon} c_m \sum_n K_e^n \psi_e^{n,in}, \\ \dfrac{\partial \psi_e^{n,in}}{\partial t} = -\dfrac{\delta^{in}}{\varepsilon} \psi_e^{n,in} - \dfrac{j_c^n\left(\Delta\psi_m + \Delta\psi_e^n\right)}{c_m K_e^n}, \\ j_e^{out}\left(\psi_m^{out}\right) = -\dfrac{\delta^{out}}{\varepsilon} c_m \sum_n K_e^n \psi_e^{n,out}, \\ \dfrac{\partial \psi_e^{n,out}}{\partial t} = -\dfrac{\delta^{out}}{\varepsilon} \psi_e^{n,out} + \dfrac{j_c^n\left(\Delta\psi_m + \Delta\psi_e^n\right)}{c_m K_e^n}, \\ \Delta\psi_m = \psi_m^{in} - \psi_m^{out}, \\ \Delta\psi_e^n = \psi_e^{n,in} - \psi_e^{n,out}, \\ n = Na, K, Cl. \end{cases} \quad (17)$$

By analogy with Eqs. 14-17 the quasi-stationary form of the equivalent electric element takes the next form:

$$\begin{cases} j_e^{in}\left(\psi_m^{in}\right) = -j_e^{out}\left(\psi_m^{out}\right) = \sum_n j_c^n\left(\Delta\psi_m + \Delta\psi_e^n\right), \\ \psi_e^{n,in} = -\dfrac{\delta^{out}}{\delta^{in}} \psi_e^{n,out} = -\dfrac{\varepsilon}{\delta^{in}} \dfrac{j_c^n\left(\Delta\psi_m + \Delta\psi_e^n\right)}{c_m K_e^n}, \\ \Delta\psi_m = \psi_m^{in} - \psi_m^{out}, \\ \Delta\psi_e^n = \psi_e^{n,in} - \psi_e^{n,out}, \\ n = Na, K, Cl. \end{cases} \quad (18)$$

Thus, Eqs. 17 are the non-stationary form and Eqs. 18 are the quasi-stationary form of the equivalent electric element of the membrane heterogeneous current. The electrical current in the equivalent element depends on the transmembrane potential as well as on intra- and extracellular conductivities of the solution.

**Cell-surface junction point-contact model**

The main interest, which remains till this point, is to figure out how the heterogeneous membrane conductivity could affect a shape of the recordable extracellular signal under various conditions. A one approach could be the development of a simplified point-contact cell-surface junction model describing the extracellular electrical arrangement between the cell and the sensor.

Sufficiently simplified point-contact model that could describe the experiment may include five compartments as referred to on the Fig. 7: a cell (*c*), an external solution (*s*), a junction between a cell and a surface (*j*), a measuring electrode (*l*) and, finally, a reference electrode (*r*).

Now in order to describe a cellular membrane the equivalent electric element of heterogeneous membrane current should be used. The non-stationary form of the equivalent electric element was exploited because of the fact that a numerical solution of Eqs. 17 is more stable. Ion currents through channels $j_C^n$ were introduced by a set of Hodgkin-Huxley equations (Hodgkin & Huxley, 1952).

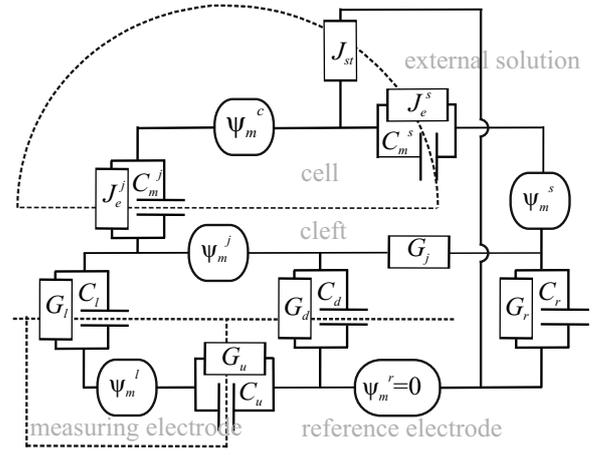

Fig. 7. Equivalent circuit of the cell-surface junction point-contact model (list of symbols in Table 1)

The Ohm's law was applied to calculate currents through other homogeneous borders. Values of parameters and sizes of boundaries are summarized in the Table 1. To calculate the seal conductance of the cleft the following formula was used (Fromherz, 2003):

$$G_j = 4\pi\delta_j h, \quad (19)$$

where, $\delta_j$ – the conductivity of the cleft, $h$ – the cleft height.

Additional transmembrane current (0.3 nA) was injected into the cell to stimulate electrical activity.

The equivalent circuit of the model corresponds to the initial value problem for a first-order differential equation system. Matlab software (MathWorks) multistep solver *ode15s* based on variable-order numerical differentiation formulas was used to solve the problem.

**Results and discussion**

Parameters, which determine a type of the cell-surface junction, are the conductivity of the cleft ($\delta_j$) and the cleft height (*h*). Both determine the seal conductance of the cleft according to the Eq. 19. At the same time, the conductivity of the cleft determines the value of the excess charge and overpotential near the membrane channels according to Eqs. 4, 6. As the consequence, the conductivity of the cleft influences the potential drop that appears across the channel, which in turn controls the channel current.

Decrease in the conductivity of the cleft results in the excess charge build up and overpotential increase near the channel pore. For sodium channels the excess charge and overpotential in the cleft have negative values. This leads to a more rapid transchannel potential depolarization (here more rapid means when compared with the membrane depolarization) and results in the early sodium channel activation. On the contrary, for potassium channels, the excess charge and overpotential in the cleft have positive values. This lowers the transchannel potential depolarization (when compared with the membrane depolarization) and reduces the channel potassium current (Fig. 8). Further decrease of the conductivity brings the potential difference over the channel down and leads to a further current recession. This effect is similar to a channel closure. As result: the less the conductivity of the cleft is, the more rapidly the sodium current increases and less of the potassium current flows (Fig. 8).

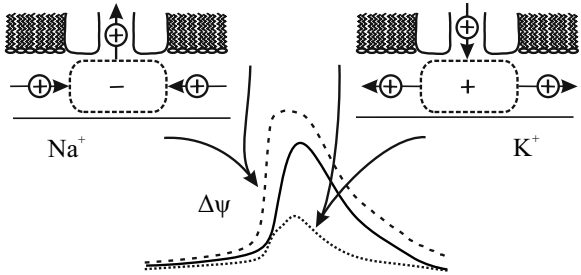

Fig. 8. Shifts in the transchannel potential waveform for sodium (--) and potassium (···) channels relative to the transmembrane potential (-) with the low conductivity in the cleft

Extracellular electrical signals simulated for different values of the conductivity of the cleft and the cleft height are shown in the Fig. 9.

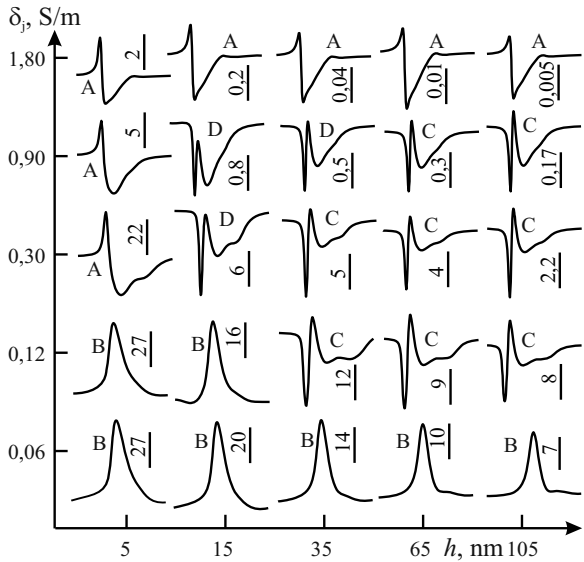

Fig. 9. Dependence of extracellular signal amplitudes (amplitude in mV, signal duration is 15 ms) and shapes (A, B, C, D-type) on the cleft height and on the conductivity of the cleft

Because of the low conductance of the measuring electrode $G_u$, its potential $\psi^j_m$ is equal to the potential in the cleft $\psi^j_m$. This electrical potential is controlled by the Kirchhoff's law, which takes the next form:

$$C_d \frac{\partial(\psi^j_m - \psi^r_m)}{\partial t} + \psi^j_m G_j = \\ = C^j_m \frac{\partial(\psi^c_m - \psi^j_m)}{\partial t} + J^j_e(\delta_j, J^j_c) \quad (20)$$

When conductivity of the cleft is large, for example equal to the conductivity of the extracellular solution, overpotential near the channels on the bottom and top membrane halves are small and equal among themselves. This situation is the symmetrical charge transfer process, when ionic and capacitive currents have similar magnitudes but opposite directions. As a result the total membrane current vanishes and the extracellular potential has small amplitude (A-type signals on Figs. 9, 10).

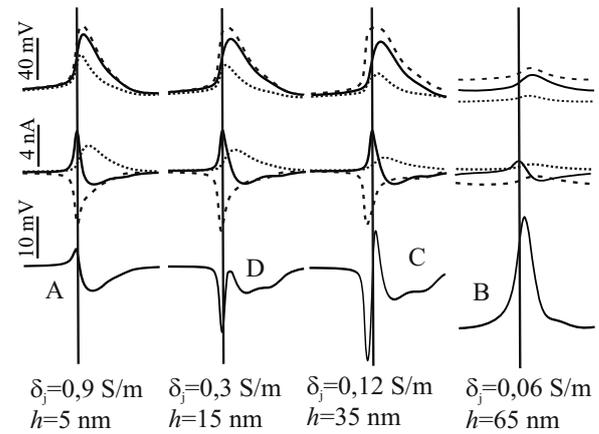

Fig. 10. Main types of extracellular electrical signals (in the third row) and corresponding to them: transchannel (-- Na, ··· K) and transmembrane (-) potentials (in the first row), currents (-- Na, ··· K, - capacitive) (in the second row), the lateral conductivity and the cleft height (in the fourth row)

Under conditions when the conductivity of the cleft is moderate and the capacitance of the substrate is small, cleft signal shape is proportional to the total membrane current and the signal amplitude depends on the seal conductance (Eq. 20). This type of contact can be called "ohmic". The total membrane current now is the sum of the current through the membrane capacitance with ionic currents through channels (Eq. 20). A more rapid sodium current increase leads to a more apparent first negative peak in the extracellular signal shape (C, D-type signals on Figs. 9, 10).

In the situation when the conductivity of the cleft is very small, say approximately thirty times less than the conductivity of the extracellular solution, the seal conductance as well as the potential difference across the channel are considerably small. If this potential falls below an excitation threshold, ionic channels of the bottom cellular membrane may not be activated. In this case the extracellular signal $\psi^j_m$ is proportional to the intracellular potential $\psi^c_m$ and the amplitude of the signal depends on the membrane and substrate capacitances (Eq. 20). This type of contact can be called "capacitive". For this contact type, the amplitude of the extracellular signal increases because of the decrease in the seal conductance (B-type signals on Figs. 9, 10).

It is interesting to note that all signals on the Fig. 9 were obtained with the same value of the seal conductance of the cleft ~54 nS (according to the Eq. 10), which corresponds to the seal resistance value of 18.5 MOhm.

Signals with shapes corresponding to the main types (A, B, C, D-type) of extracellular signals, which were found experimentally and described by other authors (Jenkner & Fromherz, 1997; Schatzthauer & Fromherz, 1998; Fromherz, 1999), could be seen among simulated extracellular signals (Figs. 9,10). Shapes of other signals represent a combination of these basic types of signals.

**Conclusion**

With the aid of the heterogeneous membrane conductivity model it was shown that changes in the passive cell-surface junction characteristic (like the conductivity of the cleft and the cleft height) may appear to be a sufficient cause of different types of extracellular signals.

Without any doubt proposed heterogeneous membrane conductivity model describes only one of possible mechanisms of the extracellular signal formation. The heterogeneous membrane conductivity mechanism was tested as alone to show its applicability in the presented point-contact model of the cell-surface junction. To describe or simulate full realistic picture of signal formation process, one have to take into consideration all possible mechanisms mentioned in the introduction.

The effects of the heterogeneous membrane conductivity will be significant if signals are registered in close cell-electrode contact. If the cells are far away from electrode, then the relative position of the cell soma and neurites will determine the signals shape (Gold *et al.*, 2006).

The point-contact model was used to simulate signals recording from current-stimulated cell. Cell stimulation can be also simulated by applying constant ore variable electric potential in one of the nodes ($\psi_m^c$, $\psi_m^j$, $\psi_m^l$, $\psi_m^s$) of the equivalent circuit of the cell-surface junction (Fig. 7).

The heterogeneous membrane conductivity model is heavily based on the continuum electrostatics to describe the charge and potential near the membrane channel. Of course, at the nanometer level Brownian and molecular dynamics methods could be preferred over the Nernst-Planck-Poisson method (Corry *et al.*, 2000). But the Nernst-Planck-Poisson theory is very useful for the ensemble-averaged description.

Hodgkin-Huxley equations, which were used to describe currents through channels (Hodgkin & Huxley, 1952) could be altered to reflect other sorts of ion channels with the current kinetics different for various types of cells. However, as a result, extracellular signal shapes could be changed in some extent.

The cleft height in the average cell-surface junction was reported to be 50-70 nm (Fromherz, 2003). More wide range of the cleft heights was intentionally used in the simulation to demonstrate the signal waveform and amplitude dependence on the height.

It is also necessary to note that the conductivity of the cleft together with the cleft height unambiguously determine electric properties of the cell-surface junction. Therefore, they can be used as the characteristic properties of the cellular adhesion to various surfaces.

Table 1 Parameters of the cell-surface junction point-contact model

| Variables and parameters | Symbol | Value |
|---|---|---|
| Intracellular potential | $\psi_m^c$ | |
| Potential in the cleft | $\psi_m^j$ | |
| Potential of the measuring electrode | $\psi_m^l$ | |
| Potential in an external solution | $\psi_m^s$ | |
| Potential of a reference electrode | $\psi_m^r$ | 0 mV |
| Dielectric permittivity of the solution | $\varepsilon$ | $81 \cdot \varepsilon_0$ F/m |
| Debye length | $\lambda$ | 1 nm |
| Cleft height | $h$ | 5÷105 nm |
| Conductivity of the cleft | $\delta_j$ | 0.06÷1.80 S/m |
| Conductivity of the external solution | $\delta_s$ | 1.8 S/m |
| Conductivity of the cell | $\delta_c$ | 0.6 S/m |
| Total membrane current of a bottom cell patch | $J_e^s$ | $j_e^s \cdot S_m^s$ |
| Total membrane current of a top cell patch | $J_e^j$ | $j_e^j \cdot S_m^j$ |
| Capacitance of the top cell patch | $C_m^s$ | $c_m \cdot S_m^s$ |
| Capacitance of the bottom cell patch | $C_m^j$ | $c_m \cdot S_m^j$ |
| Capacitance of the measuring electrode | $C_l$ | $c_l \cdot S_l$ |
| Capacitance of the reference electrode | $C_r$ | $c_r \cdot S_r$ |
| Capacitance of the measuring electrode in substrate | $C_u$ | $c_u \cdot S_u$ |
| Capacitance of the substrate | $C_d$ | $c_d \cdot S_d$ |
| Conductance of the measuring electrode | $G_e$ | $g_e \cdot S_e$ |
| Conductance of the reference electrode | $G_r$ | $g_r \cdot S_r$ |
| Conductance of the measuring electrode in substrate | $G_u$ | $g_u \cdot S_u$ |
| Conductance of the substrate | $G_d$ | $g_d \cdot S_d$ |
| Specific capacitance of the membrane | $c_m$ | 50 mF/m$^2$ |
| Specific capacitance of the measuring electrode | $c_e$ | 2 mF/m$^2$ |
| Specific capacitance of the reference electrode | $c_r$ | 2 mF/m$^2$ |

| Specific capacitance of the measuring electrode in substrate | $c_u$ | 1 mF/m$^2$ |
| --- | --- | --- |
| Specific capacitance of the substrate | $c_d$ | 3 mF/m$^2$ |
| Conductivity of the measuring electrode | $g_e$ | 1 S/m$^2$ |
| Conductivity of the reference electrode | $g_r$ | 1 S/m$^2$ |
| Conductivity of the measuring electrode in substrate | $g_u$ | 1 mS/m$^2$ |
| Conductivity of the substrate | $g_d$ | 1 mS/m$^2$ |
| Area of the top cell patch | $S_m^s$ | 2000 μm$^2$ |
| Area of the bottom cell patch | $S_m^j$ | 1000 μm$^2$ |
| Area of the measuring electrode | $S_l$ | 300 μm$^2$ |
| Area of the reference electrode | $S_r$ | 1000 mm$^2$ |
| Area of the measuring electrode in substrate | $S_u$ | 300 μm$^2$ |
| Area of the substrate | $S_d$ | 700 μm$^2$ |
| Channel density factor for sodium channels | $K_e^{Na}$ | $1 \cdot 10^5$ |
| Channel density factor for potassium channels | $K_e^K$ | $3 \cdot 10^5$ |
| Channel density factor for chlorine channels | $K_e^{Cl}$ | $2 \cdot 10^6$ |